\documentclass[%
 reprint,
 amsmath,amssymb,
 aps,
pra,
]{revtex4-2}

\usepackage{graphicx}
\usepackage{dcolumn}
\usepackage{bm}
\usepackage{xcolor}

\usepackage{amsmath,amssymb,amsfonts}
\usepackage{textcomp}
\usepackage{color,soul} 
\usepackage{xfrac}

\DeclareMathAlphabet\mathbfcal{OMS}{cmsy}{b}{n}

\begin{document}

\preprint{APS/123-QED}

\title[Breaking Reciprocity in a non-Hermitian Photonic Coupler with Saturable Absorption]{Breaking Reciprocity in a non-Hermitian Photonic Coupler with Saturable Absorption}

\author{Dimitrios Chatzidimitriou}
 \email{dchatzid@auth.gr}
 \affiliation{School of Electric and Computer Engineering, Aristotle University of Thessaloniki}
 \author{Alexandros Pitilakis}
 \affiliation{School of Electric and Computer Engineering, Aristotle University of Thessaloniki}
 \author{Traianos Yioultsis}
 \affiliation{School of Electric and Computer Engineering, Aristotle University of Thessaloniki}
\author{Emmanouil E. Kriezis}%
 \affiliation{School of Electric and Computer Engineering, Aristotle University of Thessaloniki
}%

\date{\today}

\begin{abstract}
We study the breaking of reciprocity in non-Hermitian coupled photonic waveguides by the simultaneous action of the nonlinear effect of saturable absorption, and the presence of exceptional points. The nonlinear response of such a system is studied in depth, giving physical insight into the operational principles and fundamental limitations, while producing analytical and semi-analytical results for the transmission and nonreciprocal intensity range. We numerically validate the theoretical results and assess important performance metrics for the nonreciprocal operation of such a device. This novel class of passive non-resonant devices can offer on-chip broadband nonreciprocal operation, making them potentially very important for high-speed applications.
\end{abstract}

\maketitle

\section{Introduction} \label{sec:1:intro}

Reciprocity is a fundamental electromagnetic property, imposing equal transmission coefficient when source and measurement points are interchanged. In electrical network terms, reciprocal components have a symmetric scattering matrix, $[S]=[S]^T$ \cite{Potton2004}. Reciprocity should not be confused with structural asymmetry, even though both terms are sometimes used, erroneously, in the same context \cite{Jalas2013}. The majority of passive and tunable components are reciprocal and usually exhibit a degree of structural asymmetry. Nevertheless, nonreciprocal devices are highly desirable as they are fundamental building blocks in all contemporary telecommunication and data processing systems \cite{Sounas2017}.

Nonreciprocity in optical devices is nowadays practically synonymous with the use of magneto-optical materials \cite{Dotsch2005}, which, under static magnetic biasing, exhibit non symmetric electric permittivity and magnetic permeability; these properties break classical Lorentz reciprocity and time-reversal symmetry. Unfortunately, these materials are incompatible with most PIC platforms, e.g. silicon photonics, with few exceptions \cite{Huang2016}. 

Consequently, alternative ways have been proposed to break Lorentz reciprocity on conventional low-cost PIC, such as space-time modulation \cite{Sounas2013} and nonlinearity in combination with spatial asymmetry \cite{Sounas2018,Antonellis2019}. In this work, we focus on the latter, noting that it has been mainly explored in terms of the Kerr effect (self-phase modulation) and in resonant narrow-bandwidth structures \cite{Sounas2018_IEEE_AWPL, DelBino2018,Xu2014,Rodriguez2019}. Our approach takes an alternate route, making use of the exceptional points (EPs) present in the eigenvalue space of non-Hermitian systems \cite{Miri:19}.

Non-Hermitian optics has been a very prolific field \cite{ElGanainy2019} since the initial experimental verification that the concept of Parity-Time Symmetry (PT-Symmetry) can be applied to simple coupled optical systems with exactly balanced gain and loss \cite{Rter2010}. However, PT-Symmetry is only a special case of non-Hermitian systems, meaning that similar qualitative behaviour can be attained through coupled systems with \textit{asymmetric} loss and/or gain. The EPs exhibited by such coupled optical systems have been explored in many applications such as optical switching \cite{Lupu2013,Chatzidimitriou:18}, single mode lasing \cite{Hodaei2014}, sensing \cite{Hodaei2017,Chen2017}, and, more recently, asymmetric mode switching \cite{Doppler2016,Yoon2018} or arbitrary polarization conversion \cite{Hassan2017} by encircling an EP. All these results are extremely interesting but represent reciprocal systems, as none of the aforementioned mechanism that break Lorentz reciprocity is present. Nonreciprocal optical systems with EPs have mainly been demonstrated in coupled resonators with saturable gain \cite{Chang2014}. Nonlinear assymetric directional couplers with the potential for nonreciprocal behaviour have also been investigated, but again with saturable gain and the Kerr effect as the driving nonlinear effects \cite{Kominis2016, Zhiyenbayev2019}.

Since passive optical components are in general more attractive compared to active ones, the latter being less complex and expensive to implement, in this work we will employ the nonlinear effect of saturable absorption (SA), i.e. the decrease of optical absorption with increasing incident optical intensity, as one of the most pertinent nonlinear effects to manipulate a passive non-Hermitian system. Specifically, the optical system under study is a directional coupler, a non-resonant and thus inherently broadband configuration, which consists of a lossy waveguide with saturable losses, and a lossless waveguide. It will be shown that the asymmetry imposed by the EP on the eigenmodes of the coupler, combined with the SA phenomenon, give rise to nonreciprocal response, manifested as uni-directional transmission for excitation powers in the nonreciprocial intensity range (NRIR).

The remainder of this paper is organized as follows: Section~\ref{sec:linear_coupler} discuses the operation of a linear directional coupler with asymmetric losses. Section~\ref{sec:nonlinear_coupler} includes the qualitative physical and mathematical analysis of the nonlinear nonreciprocal coupler and the derivation of the NRIR bounds. Section~\ref{sec:numerical} contains the numerical verification of the theoretical work, the numerical analysis of several key performance metrics of the nonlinear coupler, and the discussion of the results. Section~\ref{sec:5:conclusions} summarizes the conclusions of our work.

\section{A linear non-Hermitian coupler} \label{sec:linear_coupler}

In this section we will briefly review the operation of a \textit{linear} non-Hermitian photonic directional coupler and use our observations to gain physical insight into the operation of the nonlinear nonreciprocal coupler of the next section, which contains the key results of this work.

\begin{figure}[ht]
    \centering
    \includegraphics{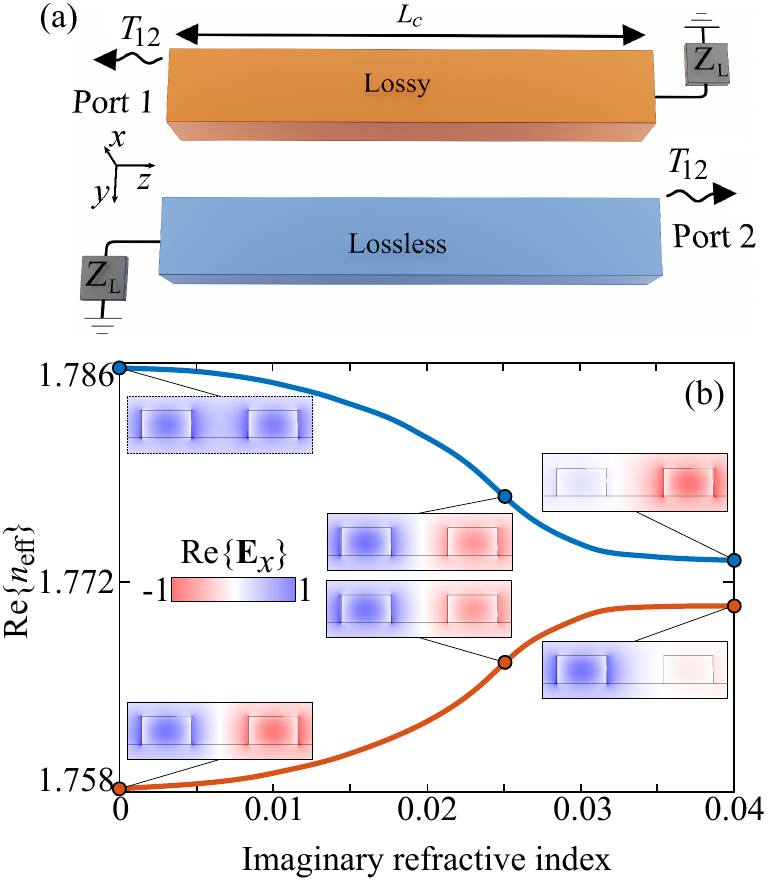}
    \caption{(a) Abstract two-port asymmetric waveguide coupler of length $L=L_c=\pi/2\kappa$. The ports that are not used are thought to be terminated to matched loads $Z_L$. $T_{21}$ and $T_{12}$ are the transmission coefficients in the forward (port 1 to 2) and backward (port 2 to 1) directions, respectively. (b) Real part of the effective refractive index of a silicon photonic coupler consisting of two identical silicon wires, as a physical example to (a). The x-axis corresponds to the artificially added imaginary part to the refractive index of one of the silicon wires. Also plotted is the real part of the $E_x$ component of the quasi-TE guided modes of the coupler at three distinct points, showing the evolution of the eigenmodes.}
    \label{fig:1}    
\end{figure}

Specifically, we consider here the abstract waveguide coupler depicted in Fig.~\ref{fig:1}, consisting of two weakly coupled single-mode waveguides: a lossy waveguide (orange), and a lossless waveguide (blue). The waveguides are otherwise identical, i.e., their individual modes have the same phase constant. The coupler is configured as a two-port component with ports 1 and 2 denoting opposite ends of the lossy and lossless waveguides, respectively. The coupler length is equal to the coupling length $L=L_c$. From now on we will refer to the transmission from port 1 to 2 ($T_{21}$) as \textit{forward} transmission and the transmission from port 2 to 1 ($T_{12}$) as \textit{backward} transmission.

A coupled mode theory (CMT) framework for the CW analysis of such a system can be developed as the following set of coupled differential equations:
\begin{equation}
\label{eq:CMT_linear_unormalized_system}
    \dfrac{d}{d z}
        \begin{bmatrix}
            B_1\\
            B_2
        \end{bmatrix}
    =
    \begin{bmatrix}
        -j\beta - {\alpha} & -j\kappa \\
        -j\kappa                             & -j\beta
    \end{bmatrix}
    \begin{bmatrix}
    B_1\\
    B_2
    \end{bmatrix}
,
\end{equation}
where $B_i$, $i=\{1,2\}$, are the complex field amplitudes of the guided modes in the $i$-th waveguide, $\beta$ is the real phase constant, $\alpha$ the loss coefficient of the mode guided in the lossy waveguide and $\kappa=\pi/2L_c$ the coupling coefficient. Note that we are using the $exp(+j\omega t)$ harmonic convention.

The system of Eq.~\eqref{eq:CMT_linear_unormalized_system} can be simplified with the following normalized parameters
\begin{subequations}
\label{eq:CMT_normalizations}
\begin{align}
z_n        &= 2\kappa z, \\ 
\alpha_n   &= \frac{\alpha}{2\kappa},\\
\beta_n    &= \frac{\beta}{2\kappa},
\end{align}
\end{subequations}
and the complex envelope transformation
\begin{equation}
\label{eq:CMT_amplitude_transformation}
    \begin{bmatrix}
            A_1\\
            A_2
        \end{bmatrix}
    =
    \begin{bmatrix}
        ~~~~B_1 \\
        -jB_2                            
    \end{bmatrix}
    e^{+j\beta_n z_n}
,
\end{equation}
which lead to the normalized equation system
\begin{equation}
\label{eq:CMT_normalized_linear_system}
    \dfrac{d}{d z_n}
        \begin{bmatrix}
            A_1\\
            A_2
        \end{bmatrix}
    =
    \begin{bmatrix}
        -{\alpha_n} & +\sfrac{1}{2} \\ 
        -\sfrac{1}{2}               & 0                 
    \end{bmatrix}
    \begin{bmatrix}
    A_1\\
    A_2
    \end{bmatrix}
.
\end{equation}
Note that the coupling length for the normalized equations is now $z_n=\pi$. {Also, since we are only studying ``half-duplex'' operation, e.g. only one port is excited at any one time, the phase difference of the complex envelopes $A_{1,2}$ can only be 0 or $\pi$. This can be proven by separating the real and imaginary parts in Eq.~\eqref{eq:CMT_normalized_linear_system}. Thus, without loss of generality, $A_{1,2}$ will be thought as real}.

The eigenvalues of Eq.~\eqref{eq:CMT_normalized_linear_system} are straightforwardly evaluated as
\begin{equation}
    \lambda_\pm = \dfrac{-\alpha_n \pm \sqrt{\alpha_n^2-1}}{2},
\end{equation}
from which we can readily identify the EP at $\alpha_n=1$, as the point in the parameter space (here the loss level) where the eigenvalues \textit{coalesce}. Note that for $\alpha_n\gg1$, $\lambda_+$ is negligible compared to $\lambda_-$. The right eigenvectors, defined as $M\mathbf{r}_\pm=\lambda_\pm\mathbf{r}_\pm$ with $M$ denoting the matrix in Eq.~\eqref{eq:CMT_normalized_linear_system}, are found as
\begin{equation}
    \label{eq:linear_right_eigenvectors}
    \mathbf{r}_\pm = \dfrac{1}{\sqrt{2}}
    \begin{bmatrix}
    { \sqrt{1\mp \sqrt{1-1/\alpha_n^2}}} \\ \\
   {\sqrt{1\pm\sqrt{1-1/\alpha_n^2}}}
    \end{bmatrix},
\end{equation}
where the normalization was done assuming $\alpha_n\geq1$. The $\mathbf{r}_\pm$ correspond to low and high losses, respectively.

The \textit{coalescence} term is used to signify that not only the eigenvalues but also the respective eigenvectors converge at the EP, as can be seen from Eq.~\eqref{eq:linear_right_eigenvectors}. We believe it is also helpful to showcase this behaviour in a physical setting. Thus, as an example, we employ a silicon photonic coupler operating at a wavelength of $1.55~\mathrm{\mu m}$. The coupler consists of two silicon wires ($350~\mathrm{nm}$ wide and $200~\mathrm{nm}$ tall) with refractive index $n_\mathrm{Si}=3.47$ separated by a $400~\mathrm{nm}$ wide gap. The silicon wires are air-clad and are set on a silicon oxide substrate ($n_\mathrm{SiO_2}=1.45$). We now artificially add an imaginary part to the refractive index of one of the waveguides, denoting that one waveguide is lossy whereas the other is lossless. The effective refractive index ($n_\mathrm{eff}$) as well as the cross-section field profiles of the TE-polarized guided modes are numerically evaluated through a finite element method (FEM) modal analysis of the waveguide and plotted in Fig~\ref{fig:1}(b). {Note that the blue and red curves do not actually touch. Strictly speaking, that means that the modes do not reach the EP, since the change of the imaginary part of the refractive index also changes the real part of the phase constant and thus the modes do not perfectly coalesce \cite{Benisty:12}.}

Let us now discuss the effect that the presence of the EP has on the eigenmodes of a lossy photonic coupler. When losses are very low (we will refer to this as ``operation below the EP'') the eigenmodes resemble that of a conventional low-loss coupler, the symmetric and anti-symmetric supermode, with field profiles equivalently distributed across both waveguides. This is the leftmost set of mode profiles shown in Fig.~\ref{fig:1}(b). As losses increase, the eigenmodes become less and less orthogonal to each other (as per the conjugated orthogonality condition) and eventually coalesce to the singularity of the EP. This is represented by the middle set of mode profiles in Fig.~\ref{fig:1}(b), where as discussed the eigenmodes do not perfectly coalesce but are instead located very close the EP. As losses increase further (we will refer to this as ``operation above the EP'') the eigenmode field distributions asymptotically approach the modes of the isolated waveguides, as if the latter were not coupled [rightmost set of mode profiles in Fig.~\ref{fig:1}(b)]. Thus, when operating above the EP the two eignemodes have a different overlap with the material of each waveguide. In this example, since one waveguide is lossy and the other lossless, the eigenmode associated with the lossy waveguide will be attenuated much faster than the other along the propagation.

Returning to the normalized set of Eqs.~\eqref{eq:CMT_normalized_linear_system} we can analytically solve the system to study how light propagates in such a system. For $\alpha_n\ll1$ light is almost fully coupled from one waveguide to the other at the coupling length, as per the conventional directional coupler operation, which is extensively covered in literature. Thus, we are mostly interested in the $\alpha_n\gg1$ regime, which can also be found in the literature but is not as trivial. We present below the solution to Eq.~\eqref{eq:CMT_normalized_linear_system} with initial conditions $A_1(0)$ and $A_2(0)=0$ (excitation of the lossy waveguide)
\begin{subequations}
    \label{eq:CMT_Forward_transmission_LinearAnalytic_solution}
    \begin{align}
    \label{eq:CMT_Forward_transmission_LinearAnalytic_solution_a}
    A_1(z_n) &\ = \frac{A_1(0)}{\sqrt{\alpha_n^2-1}} \left(\lambda_{+}e^{\lambda_{+}z_n} -\lambda_{-}e^{\lambda_{-}z_n}\right), \\
    \label{eq:CMT_Forward_transmission_LinearAnalytic_solution_b}
    A_2(z_n) &\ = \frac{A_1(0)}{2\sqrt{\alpha_n^2-1}} \left( -e^{\lambda_{+}z_n} + e^{\lambda_{-}z_n}\right),
    \end{align}
\end{subequations}
and also for initial conditions $A_1(0)=0$ and $A_2(0)$ (excitation of the lossless waveguide)
\begin{subequations}
    \label{eq:CMT_Backward_transmission_LinearAnalytic_solution}
    \begin{align}
    \label{eq:CMT_Backward_transmission_LinearAnalytic_solution_a}
    A_1(z_n) &\ = \frac{A_2(0)}{2\sqrt{\alpha_n^2-1}} \left( e^{\lambda_{+}z_n} - e^{\lambda_{-}z_n}\right), \\
    \label{eq:CMT_Backward_transmission_LinearAnalytic_solution_b}
    A_2(z_n) &\ = \frac{A_2(0)}{\sqrt{\alpha_n^2-1}} \left(-\lambda_{-}e^{\lambda_{+}z_n} +\lambda_{+}e^{\lambda_{-}z_n}\right).
    \end{align}
\end{subequations}
The exponential terms $\exp{(\lambda_{\pm}z_n)}$ can be linked to the eigenmodes of the physical example presented in Fig.~\ref{fig:1}(b), with $\exp{(\lambda_{+}z_n)}$ corresponding to the low attenuation eigenmode propagating mainly in the lossless waveguide, and $\exp{(\lambda_{-}z_n)}$ corresponding to the highly attenuated eigenmode confided mainly inside the lossy waveguide. Since $|\lambda_-|\gg|\lambda_+|$, exciting the lossy/lossless waveguide corresponds to primarily exciting the eigenmode that travels in the respective waveguide [see Eq.~\eqref{eq:CMT_Forward_transmission_LinearAnalytic_solution_a} and Eq.~\eqref{eq:CMT_Backward_transmission_LinearAnalytic_solution_b}]. Thus, in the forward direction light will be greatly attenuated while in the backward direction light will tend to stay in the lossless waveguide. Note though, that the linear system is completely reciprocal; asymmetry alone cannot break reciprocity.

\begin{figure}[t]
    \centering
    \includegraphics{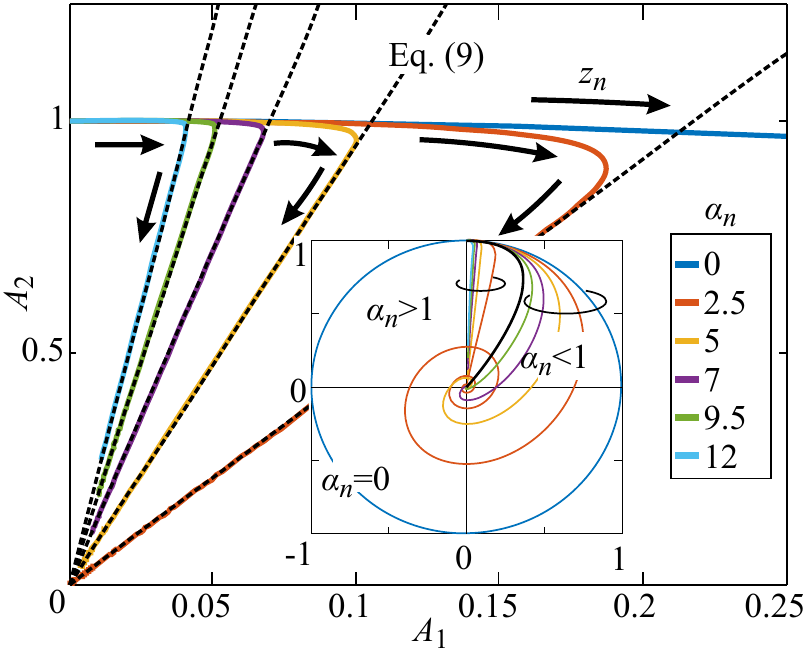}
    \caption{(a) Parametric solutions $A_1(z_n)$ and $A_2(z_n)$ of the linear system of Eq.~\eqref{eq:CMT_normalized_linear_system} for various values of $\alpha_n$. The inset shows the general evolution of the trajectories grouped by whether or not the system is above ($\alpha_n>1$) or below ($\alpha_n<1$) the EP. The black trajectory is exactly on the EP, $\alpha_n=1$.}
    \label{fig:2}    
\end{figure}

Finally, focusing on the backward propagation (excitation from the lossless coupler), we plot in Fig.~\ref{fig:2} the parametric curves defined by Eq.~\eqref{eq:CMT_Backward_transmission_LinearAnalytic_solution}, i.e., the field amplitude trajectories along the coupler, for various values of $\alpha_n$ and for initial conditions $A_2(0)=1$ and $A_1(0)=0$. The inset shows a more general picture of the trajectories including the regime $\alpha_n<1$. For $\alpha_n=0$ the trajectory is a circle and for $0<\alpha_n<1$ the trajectories are spirals towards the origin. For $\alpha_n>1$, the most interesting feature we observe is that all trajectories converge to straight lines, or equivalently that $A_2(z_n)/A_1(z_n)=\mathrm{const.}$ for large enough $z_n$. This is understood as the regime where the high-loss eigenmode/eigenvector vanishes. 

This can also be shown from the analytical solution, since the $\exp(\lambda_-z_n)$ term goes to zero a lot faster than $\exp(\lambda_+z_n)$ so that Eq.~\eqref{eq:CMT_Backward_transmission_LinearAnalytic_solution} becomes
\begin{equation}
\label{eq:CMT_Backward_transmission_linear_convergence}
\dfrac{A_2(z_n)}{A_1(z_n)}\approx -2\lambda_-=\alpha_n+\sqrt{\alpha_n^2-1},
\end{equation}
which is exactly the direction of the $\mathbf{r}_+$ eigenvector. In Fig.~\ref{fig:2}, Eq.~\eqref{eq:CMT_Backward_transmission_linear_convergence} is depicted with dashed black lines.

\section{Non-Hermitian coupler with SA} \label{sec:nonlinear_coupler}
In this section we will describe the nonreciprocal operation of the nonlinear asymmetric coupler. The setup is the same as the coupler in Fig.~\ref{fig:1}(a) but this time the first waveguide exhibits the phenomenon of SA instead of constant linear losses, i.e., light absorption decreases with increasing optical power ($|A_1|^2$).

Assuming SA follows a simple phenomenological law, we can write the coupled mode equations as
\begin{equation}
\label{eq:CMT_nonlinear_system}
    \dfrac{d}{d z}
        \begin{bmatrix}
            B_1\\
            B_2
        \end{bmatrix}
    =
    \begin{bmatrix}
        -j\beta - \dfrac{\alpha}{1+|B_1|^2/P_\mathrm{sat}} & -j\kappa \\
        -j\kappa                             & -j\beta
    \end{bmatrix}
    \begin{bmatrix}
    B_1\\
    B_2
    \end{bmatrix}
,
\end{equation}
where ${1}/(1+|B_1|^2/P_\mathrm{sat})$ is the SA saturation function, $P_\mathrm{sat}$ the saturation power (defined as the CW power for which losses are halved), and $\alpha$ the low-power attenuation coefficient for the lossy waveguide. The system of Eq.~\eqref{eq:CMT_nonlinear_system} can be simplified under the same normalizations that were used in the linear case with the additional normalization of the $B_{1,2}$ amplitudes to the square root of the saturation power $P_\mathrm{sat}$. The normalised system is then written as
\begin{equation}
\label{eq:CMT_normalized_nonlinear_system}
    \dfrac{d}{d z_n}
        \begin{bmatrix}
            A_1\\
            A_2
        \end{bmatrix}
    =
    \begin{bmatrix}
        -\dfrac{\alpha_n}{1+|A_1|^2} & +\sfrac{1}{2} \\ 
        -\sfrac{1}{2}               & 0                 
    \end{bmatrix}
    \begin{bmatrix}
    A_1\\
    A_2
    \end{bmatrix}
.
\end{equation}
Throughout this section we will assume that $\alpha_n\gg1$, which means that for very low $A_1$ the operation of the coupler is set above the EP, as was discussed in the previous section. The instantaneous eigenvalues of Eq.~\eqref{eq:CMT_normalized_nonlinear_system} are evaluated as $\lambda_\pm(z_{n}) = [-\alpha_n/(1+|A_1|^2)\pm \sqrt{\alpha_n^2/(1+|A_1|^2)^2-1}]/2$, from which we can identify that the EP is now power dependent. Specifically, the instantaneous eigenvalues/eigenvectors coalesce when $|A_1|^2 = \alpha_n-1$.

Based on the physical intuition gained from studying the linear system we can qualitatively understand and predict how such an optical system operates. For excitation from the lossy waveguide (the forward direction), if power is high enough to saturate losses below the EP, light will start coupling to the lossless waveguide. As light leaves the nonlinear waveguide, its losses will start to desaturate, the eigenmodes will approach the EP, and consequently coupling will be decreased. When there is very little light left in the nonlinear waveguide (from the combination of coupling and absorption) the operation is high above the EP, so the lossy eigenmode will quickly vanish and the light propagating in the lossless waveguide will continue traveling in the lossless waveguide until it reaches the exit port. Consequently, the transmission in the forward direction is high when an acceptable portion of the initial excitation has coupled to the lossless waveguide \textit{before} operation shifts above the EP. 

In the backward direction, the excitation from the lossless waveguide has no impact on the eigenmodes of the coupler since there is no initial interaction with the nonlinear waveguide. Thus, only a very small fraction of the injected power will be able to couple to the nonlinear waveguide, since the eigenmode confined in the nonlinear waveguide is very weakly excited. We can distinguish three different cases depending on the power that is coupled to the nonlinear waveguide: In the first case, power does not sufficiently saturate losses and operation remains above the EP, so that light in the lossy waveguide will be quickly attenuated and the remaining light in the lossless waveguide will keep propagating in the lossless waveguide. In the second case, the system shifts at first below the EP but without saturating losses enough, so that power in the nonlinear waveguide begins decreasing and consequently the system shifts back above the EP, prohibiting further coupling. Note the difference with the forward direction, where light that was coupled to the opposite waveguide always reaches the exit port as it was ``stored'' in the low-loss eigenmode. Another important aspect to underline is that the more power is coupled to the nonlinear waveguide the easier it gets for light to continue coupling, thus potentially creating an avalanche effect for sufficiently high excitation. In the third case at very high input powers, if the power coupled to the nonlinear waveguide saturates losses so as to shift operation below the EP for the majority of propagation, then light will exit through the lossy waveguide with high transmission. 

To summarize, for sufficiently \textit{high} excitation, propagation in the forward direction has high overlap with the nonlinear material which leads to high transmission. For sufficiently \textit{low} excitation, propagation in the backward direction has low overlap with the nonlinear material and leads to low transmission. For \textit{very low} initial excitation the coupler behaves as a high-loss linear coupler with very low transmission in both directions. For \textit{very high} initial excitation the coupler behaves as a low-loss coupler with high transmission in both directions.

The guided power thresholds that define the boundaries of the regime where we expect high transmission from port 1 and low transmission from port 2, define the NRIR. Note that whereas in most common applications of nonlinear nonreciprocity, for example in a Kerr resonator \cite{Sounas:2018}, the system is only perturbed by nonlinearity, in this work the initial conditions drastically transform the eigenmodes of the coupler.

We underline that the existence of the EP enhances the asymmetry of the system but does not break reciprocity without nonlinearity. This is evident by the analysis of the linear system. Finally, note that the choice of ports 1 and 2 as the opposite ends of different waveguides is crucial to render the system asymmetric.

In the following subsections we will look more closely to the forward and backward propagation cases as well as produce estimates for the NRIR of the device.

\subsection{Forward Transmission\label{sec:Forward_Transmission}}

In the forward operation light is injected from the lossy waveguide [$A_2(0)=0$] and ideally exits with high transmission from the lossless waveguide. This scenario can be attained when initial power is high enough to ensure that for the majority of propagation losses are saturated sufficiently to operate below the EP and also ensure acceptable ILs. This means that in an \textit{averaged sense} the coupler will resemble a low-loss linear coupler.

Thus, we will approximate the solution of the nonlinear system of Eq.~\eqref{eq:CMT_normalized_nonlinear_system} with a linear system with constant losses set by the initial excitation such as

\begin{figure}[t]
    \centering
    \includegraphics{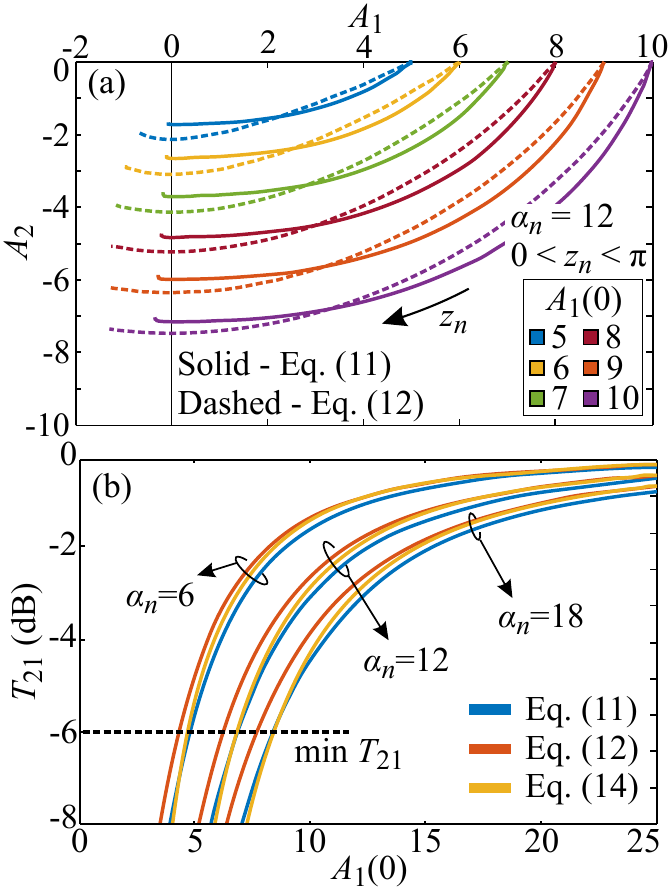}
    \caption{(a) Parametric solutions $A_1(z_{n})$ and $A_2(z_{n})$ in the forward propagation direction for various initial excitations. Solid curves correspond to numerically evaluated solutions, while dashed curves to the linear approximation. (b) Comparison of the numerical and linearly approximated power transmission coefficient in the forward propagation direction versus input amplitude. The horizontal dashed line corresponds to the minimum acceptable transmission coefficient in the forward direction for $z_{n}=\pi$.}
    \label{fig:forward_transmission_approx}    
\end{figure}

\begin{equation}
\label{eq:CMT_forward_transmission_linear_system_approx}
    \frac{d}{d z_n}
        \begin{bmatrix}
            A_1\\
            A_2
        \end{bmatrix}
    =
    \begin{bmatrix}
        -\dfrac{\alpha_n}{1+|hA_1(0)|^2}    & +\sfrac{1}{2} \\ 
        -\sfrac{1}{2}  & 0
    \end{bmatrix}
    \begin{bmatrix}
    A_1\\
    A_2
    \end{bmatrix}
,
\end{equation}
where $h<1$ is a dimensionless parameter, introduced to take into account the fact that as the initial power in the lossy waveguide 1 is inevitably coupled to the lossless waveguide 2, absorption will de-saturate and the remaining light still propagating in the lossy waveguide will experience increased losses. This is why we stated that the nonlinear coupler will resemble a low-loss coupler in an averaged sense. In Fig.~\ref{fig:forward_transmission_approx} we compare the numerical solution of Eq.~\eqref{eq:CMT_normalized_nonlinear_system} to analytical solutions of Eq.~\eqref{eq:CMT_forward_transmission_linear_system_approx} for various initial conditions $A_1(0)$ and find that the linear system, fitted with parameter $h=0.75$, provides satisfactory agreement. The solutions are presented in the $(A_1,A_2)$ plane as parametric curves $A_1 = A_1(z_n), A_2=A_2(z_n)$ and the direction of the curve is denoted with an arrow with the annotation $z_n$.
From the solution of Eq.~\eqref{eq:CMT_forward_transmission_linear_system_approx} and for initial conditions $A_1(0)$ and $A_2(0)=0$ we extract the transmission equation
\begin{subequations}
\label{eq:CMT_forward_transmission_approx}
\begin{align}
\label{eq:CMT_forward_transmission_approx_a}
\left|\frac{A_2(\pi)}{A_1(0)}\right| &= \frac{1}{w}\sin\left(w\frac{\pi}{2}\right)e^{-\frac{\alpha_n}{1+|h A_1 (0)|^2}\frac{\pi}{2} }, \\
\label{eq:CMT_forward_transmission_approx_b}
w&=\sqrt{1-\left[\frac{\alpha_n}{1+|h A_1(0)|^2 }\right]^2 }.
\end{align}
\end{subequations}

For transmission to be high in the above equation $w\approx1$ or equivalently $\alpha_n/[1+|hA_1(0)|^2]\ll 1$. Thus, we can simplify Eq.~\eqref{eq:CMT_forward_transmission_approx_a} as
\begin{equation}
\label{eq:CMT_forward_transmission_T(A1)_approx}
    T_{21}(\pi)=\left|\frac{A_2(\pi)}{A_1(0)}\right|=e^{-\frac{\alpha_n}{1+|h A_1 (0)|^2}\frac{\pi}{2} },
\end{equation}
which can be solved for $A_1(0)$ leading to
\begin{equation}
\label{eq:CMT_forward_transmission_A1(T)_approx}
    |A_1(0)|^2 = -\frac{1}{h^2} \left( \frac{\pi\alpha_n}{\ln{|T_{21}(\pi)|^2}} + 1 \right)
\end{equation}
So, given a desired minimum forward transmission $T_\mathrm{min}=T_{21}(\pi)$ and Eq.~\eqref{eq:CMT_forward_transmission_A1(T)_approx} we can approximate the minimum input power required to achieve $T_\mathrm{min}$. Thus, Eq.~\eqref{eq:CMT_forward_transmission_A1(T)_approx} stands as the lower power limit for operation in the forward direction. In Fig.~\ref{fig:forward_transmission_approx}(b) we compare the power transmission as calculated by Eq.~\eqref{eq:CMT_normalized_nonlinear_system}, \eqref{eq:CMT_forward_transmission_linear_system_approx} and \eqref{eq:CMT_forward_transmission_T(A1)_approx} for a range of initial input powers. It can be clearly seen that the approximation is very satisfactory.

Looking at Eq.~\eqref{eq:CMT_forward_transmission_A1(T)_approx} we can make two qualitative observations. First, as expected, the higher $T_\mathrm{min}$ demanded the higher input power is required. Secondly, the power threshold $|A_1 (0)|^2$ is also a strictly increasing function of $\alpha_n$.

\subsection{Backward Transmission\label{sec:CMT_Backward_Transmission}}
In the backward operation the coupler is excited from the lossless waveguide [port 2, in the right side of Fig.~\ref{fig:1}(a)], which is equivalent to initial conditions $A_2(0)$ and $A_1(0)=0$. Here, we will try to determine the upper bound of the NRIR, i.e., the maximum backward excitation for which backward transmission is sufficiently low.

Because the upper bound of the NRIR is between the regimes of low and high transmission, we will not attempt to find a linear approximation as we did in the forward direction. Instead, we will identify the limit with regard to the power flow into the nonlinear waveguide. To do so, it is very helpful (as a visual aid) to define the position vector $\mathbf{A}(z_n)= [A_1(z_n),A_2(z_n)]$ and the velocity vector field $\mathbf{v}=d\mathbf{A}/dz_n$ from Eq.~\eqref{eq:CMT_normalized_nonlinear_system}. 

\begin{figure}[t]
    \centering
    \includegraphics{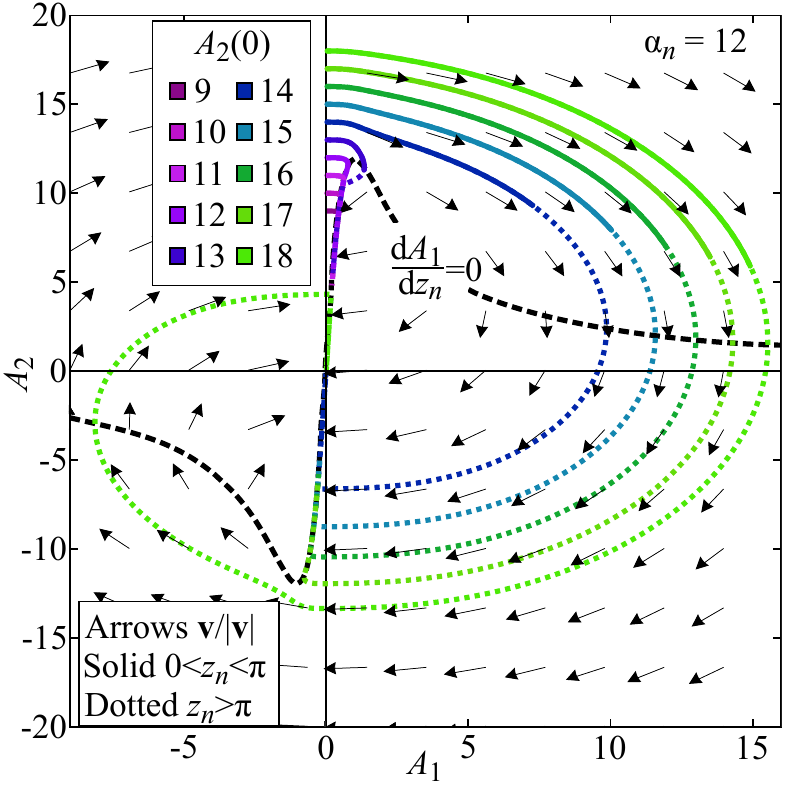}
    \caption{Velocity vector field of Eq.~\eqref{eq:CMT_normalized_nonlinear_system} for $\alpha_{n}=12$. The colored curves plotted correspond to numerically evaluated parametric solutions $A_1(z_{n})$ and $A_2(z_{n})$ for the respective initial condition. The dashed black curve corresponds to Eq.~\eqref{eq:CMT_da1=0}, i.e. $dA_1/dz_n=0$.}
    \label{fig:Velocity_vector_field}    
\end{figure}

The normalized velocity field $\mathbf{v}/|\mathbf{v}|$ is plotted in Fig.~\ref{fig:Velocity_vector_field}, alongside a number of numerically evaluated trajectories (colored solid curves). It can be intuitively understood [or shown from Eq.~\eqref{eq:CMT_normalized_nonlinear_system}] that the only fixed point [i.e., the only location in the $(A_1,A_2)$-plane where $\mathbf{v}=\mathbf{0}$] in this passive absorbing system is the origin, denoting that all light has been absorbed. From the shape of the velocity vector field, the qualitative long-term evolution of the trajectories is that they spiral inward and after sufficient propagation distance they asymptotically converge on the same path towards the origin, $A_{1,2}=0$. The path along which the trajectories converge is identified as
\begin{equation}
    \label{eq:CMT_Backwards_transmission_nonLinear_convergence}
    \dfrac{A_2}{A_1}\approx-2\lambda_{-}(z_{n})=\left[\dfrac{\alpha_n}{1+|A_1|^2} + \sqrt{\left(\dfrac{\alpha_n}{1+|A_1|^2}\right)^2-1}\right],
\end{equation}
which is the direction of the instantaneous low-loss eigenvector as it reaches its limit at $A_1=0$. This confirms the qualitative discussion at the start of the section where we argued that as long as there is low power propagating in the nonlinear waveguide, then saturation of losses will be relatively weak and thus one of the eigenmodes will vanish much faster than the other.

The dashed black line in Fig.~\ref{fig:Velocity_vector_field} represents the geometric locus where $dA_1^2/dz_{n} = 2A_1dA_1/dz_{n}=0$ which is found to be
\begin{equation}
    \label{eq:CMT_da1=0}
    A_2 = \dfrac{2\alpha_n}{1+|A_1|^2}A_1.
\end{equation}
Geometrically, Eq.~\eqref{eq:CMT_da1=0} means that the velocity is at an angle of $-\pi/2$ (points downward) and, physically, that all the power coupled to the lossy waveguide is attenuated.

In order to estimate a power threshold for low transmission in the backwards direction we need to find an initial condition that guarantees the solution does not stop at a high $A_1$ for $z_{n}=\pi$. Unfortunately, since we cannot analytically solve the nonlinear system we cannot make predictions for a specific $z_{n}$. One could try to find an approximate solution with perturbation theory by constructing a power series, which can be the focus of future investigation, but in this work we will follow a simpler approach: For $z_n\leq\pi$, when a solution satisfies Eq.~\eqref{eq:CMT_da1=0} for small values of $A_1/A_2(0)$, transmission will be low even if propagation continues, since from that point on guided power in the lossy waveguide is decreasing ($dA_1/dz_n<0$). Thus, we find the maximum $A_2$ so that $dA_1/dz_n=0$. From Eq.~\eqref{eq:CMT_da1=0} this is found to be $A_{2}=\alpha_n$ at $A_1=1$. 

We now need to translate the point $(A_1,A_2)=(1,\alpha_n)$ to an initial condition, i.e., trace it back to $z_n=0$ to find $A_2(0)$. We do so under the following reasoning: If there were no losses in the system the initial excitation would be equal to $\sqrt{\alpha_n^2+1}$. Since there are losses (solutions spiral inward), it is guaranteed that $A_2(0)<\sqrt{\alpha_n^2+1}$ will lead to solutions that satisfy Eq.~\eqref{eq:CMT_da1=0} for $A_1<1$. So all excitations up to $\sqrt{\alpha_n^2+1}$ can be assumed to lead to low transmission and consequently the backward propagation power threshold is given by
\begin{equation}
    \label{eq:CMT_backward_propagation_threshold}
    A_2(0)\leq\sqrt{\alpha_n^2+1}.
\end{equation}
Note that this limit is not absolute since higher initial excitations will still lead to low transmission. We intuitively expect though that this threshold will provide a better prediction for very low transmission as $\alpha_n$ increases. Specifically when $\alpha_n/(1+|A_1|^2)\gg1$ for $A_1\leq1$ the eigenmodes of the coupler will have a very high contrast with regards to losses. Taking into account that the higher the power propagating in the nonlinear waveguide the easier the coupling of light will be, we expect that the transmission threshold will get sharper as the distance between the eigenvalues $|\lambda_+-\lambda_-|$ at $A_1=1$ increases.

We can now estimate the transmission by considering the following inequalities
\begin{subequations}
\begin{align}
    \label{eq:backward_transmission_inequalities}
    A_1(\pi)<A'_1 \\
    A_2(0)>A'_2
\end{align}
\end{subequations}
where we use the prime notation for the value that $A_{1,2}$ have when they satisfy Eq.~\eqref{eq:CMT_da1=0}. From Eq.~\eqref{eq:backward_transmission_inequalities} and for $A_1\leq1$, we can readily construct the inequality
\begin{equation}
    T_{12}=\left|\dfrac{A_1(\pi)}{A_2(0)}\right|< \left|\dfrac{A_1'}{A_2'}\right|=\dfrac{1+|A'_1|^2}{2\alpha_n}<\dfrac{1}{\alpha_n},
\end{equation} 
{which, under the assumptions and approximations made, shows the remarkably simple limit}
\begin{equation}
\label{eq:CMT_backwards_transmission_T_bound_2}
    T_{12} < \dfrac{1}{\alpha_n}.
\end{equation}

\subsection{NRIR calculation}

We combine the forward power threshold from Eq.~\eqref{eq:CMT_forward_transmission_A1(T)_approx} with the backward power transmission threshold from Eq.~\eqref{eq:CMT_backward_propagation_threshold} to estimate the NRIR as the ratio of the forward to backward transmission
\begin{align}
    \label{eq:CMT_NRIR_estimate_normalized}
    \mathrm{NRIR} = \dfrac{\alpha_{n}^2+1}{-\dfrac{1}{h^2} \left( \dfrac{\pi\alpha_n}{\ln{|T_\mathrm{min}|^2}} + 1\right)} .
\end{align}
Keep in mind that this formula sets the backward power transmission limit to less than $1/\alpha_n^2$ and not a constant value. Nevertheless, under the $\alpha_n \gg 1$ condition, it translates to sufficiently high isolation (low transmission) in the backward direction.

There are a number of important observations that can be made from Eq.~\eqref{eq:CMT_NRIR_estimate_normalized}. First of all, NRIR is independent of the saturation power $P_\mathrm{sat}$. Secondly, increasing the forward transmission $T_\mathrm{min}$ reduces the NRIR, and vice versa. Thirdly, NRIR scales with $\alpha_n$ which offers an alternate route to increase the NRIR besides lowering $T_\mathrm{min}$. This is in contrast to nonlinear nonreciprocal devices with a single resonator, where the only way to increase the NRIR is to reduce transmission. We underline that increasing the critical parameter $\alpha_n=\alpha L_c/\pi$ can be achieved either by increasing the low-power losses ($\alpha$) through waveguide/material engineering, or by increasing the coupling length ($L_c$) through the geometric distance of the waveguides of the coupler. Thus, for any $T_\mathrm{min}< 1$ we can always choose $\alpha_n$ high enough to keep the NRIR constant. Of course, for very high $T_\mathrm{min}$, the parameters $\alpha$ and $L_c$ would take unrealistic values. Finally, for $T_\mathrm{min}=1$ there can be no NRIR window, implying that this kind of system can never have nonreciprocal behaviour with perfect transmission. This restriction is physically consistent with our analysis since $|T_\mathrm{min}|=1$ only happens for infinite input power, which would reduce losses to zero and thus make the coupler reciprocal.

\section{Numerical verification, analysis and discussion} \label{sec:numerical}
In this section we will numerically verify the theorical boundaries of the NRIR extracted in Section~\ref{sec:nonlinear_coupler} and provide further operation metrics of the nonreciprocal coupler.

\begin{figure}
    \centering
    \includegraphics[width=8.5cm]{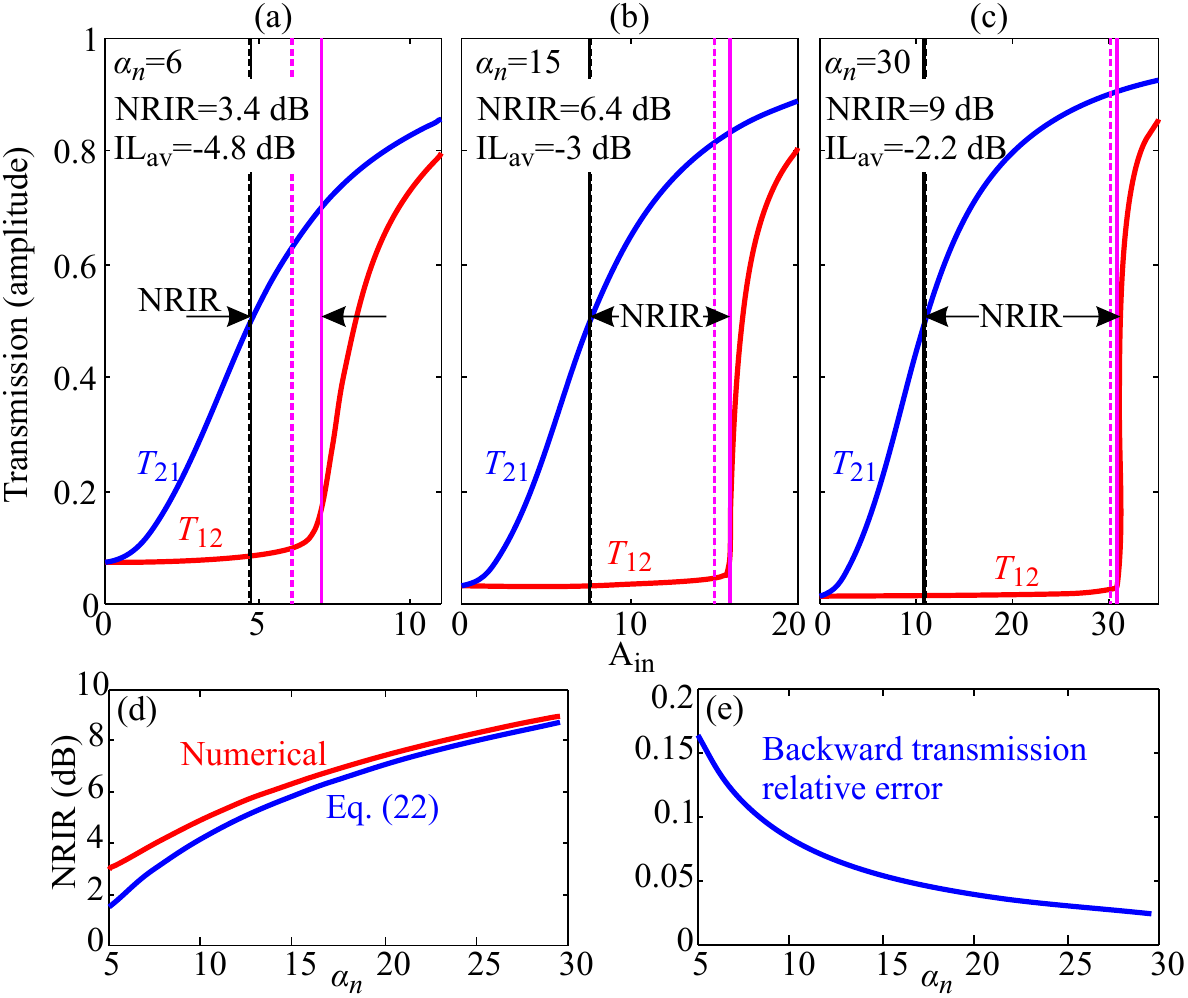}
    \caption{(a)-(c) Numerically evaluated transmission coefficient in the forward ($T_{21}$) and backward direction ($T_{12}$) direction for $\alpha_n=6,15,30$, respectively. Black arrows denote the NRIR. Black solid and dashed curves denote the numerically and theoretically [from Eq.~\eqref{eq:CMT_forward_transmission_A1(T)_approx}] evaluated lower bound of the NRIR, respectively. Magenta solid and dashed curves denote the numerically and theoretically [from Eq.~\eqref{eq:CMT_backward_propagation_threshold}] evaluated upper bound of the NRIR, respectively. Also noted are the forward ILs averaged inside the numerically evaluated NRIR. (d) Comparison of the numerically and theoretically evaluated NRIR versus $\alpha_n$. (e) Relative error between the numerically and theoretically evaluated input amplitude at the upper bound of the NRIR (backward transmission threshold).}
    \label{fig:anal_vs_nume}
\end{figure}

Using Eq.~\eqref{eq:CMT_normalized_nonlinear_system} for $\alpha_n=6,15,~\mathrm{and}~30$ we numerically evaluate and plot in Fig.~\ref{fig:anal_vs_nume}(a)-(c) the forward ($T_{21}$) and backward ($T_{12}$) transmission coefficient versus the input amplitude $A_\mathrm{in}$. The NRIR is marked with black arrows and is found from the numerical curves as the input range that leads to $T_{21}>-6~\mathrm{dB}$ and $T_{12}<1/\alpha_n$. Note that the backward transmission threshold is not constant between (a)-(c) so that we can correctly compare the numerically evaluated limits with those predicted by Eq.~\eqref{eq:CMT_backward_propagation_threshold} and \eqref{eq:CMT_backwards_transmission_T_bound_2}. Black and magenta vertical lines correspond to the forward and backward transmission thresholds, respectively. The horizontal lines are further distinguished by being solid or dashed, referring to numerical or theoretical evaluation, respectively. The black dashed/solid vertical lines are practically identical, since Eq.~\eqref{eq:CMT_forward_transmission_A1(T)_approx} was already fitted to numerically evaluated solutions with parameter $h$. Nevertheless, it is interesting that a constant value of $h=0.75$ provides such an accurate prediction over a huge range of $\alpha_n$. Comparing the magenta solid/dashed vertical lines we can see that as $\alpha_n$ increases the theoretical prediction becomes very accurate. This is more clearly shown in Fig.~\ref{fig:anal_vs_nume}(d) and (e) where we plot the NRIR versus $\alpha_n$ as well the relative error between the numerically evaluated and theoretically predicted backward transmission threshold. This behaviour is compatible with the analysis and discussion of the previous section, since the theoretical limits were extracted for $\alpha_n/(1+|A_1|^2)\gg1$ around $A_1=1$. 

\begin{figure}
    \centering
    \includegraphics{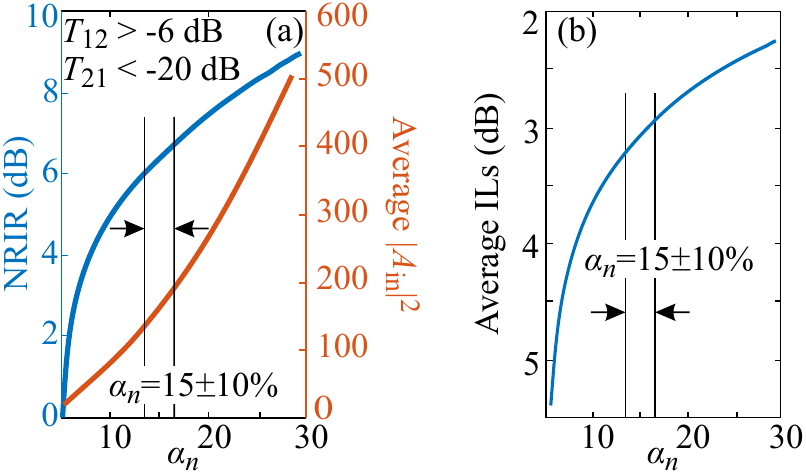}
    \caption{(a) Numerically evaluated NRIR (blue) and average input power in the NRIR window (orange) versus $\alpha_n$. (b) Average ILs in the NRIR window versus $\alpha_n$. In both (a), (b) the vertical black lines and arrows denote the change of the performance metrics when $\alpha_n=15$ is varied by $\pm10\%$ due to dispersion effects or fabrication deviations.}
    \label{fig:numerical}
\end{figure}

In the previous paragraph the backward transmission threshold of the NRIR was not constant, but varied as a function of $\alpha_n$. This was done for the sake of numerically verifying the theoretical results. However, in order to comparatively assess the performance of such a nonreciprocal device the bounds of the NRIR should be constant. Consequently, we now choose the limits of the NRIR as $T_{21}>-6~\mathrm{dB}$ (same as before) and $T_{12}<-20~\mathrm{dB}$. Fig.~\ref{fig:numerical}(a) shows some of the most important metrics in nonlinear nonreciprocal devices versus $\alpha_n$: the NRIR and the central input power for operation in the NRIR, the latter being the average of the input power at the bounds of the NRIR. Fig.~\ref{fig:numerical}(b) supplements the previous metrics with the average ILs for operation in the NRIR. Compared to resonant nonlinear nonreciprocal devices, ILs are higher but are not prohibitive. It is important to note though, that it is possible to simultaneously increase the NRIR and decrease the average ILs by increasing $\alpha_n$, contrary to nonlinear nonreciprocal devices with a single resonator \cite{Sounas:2018}, where in order to increase NRIR you have to decrease forward transmission. Recently this restriction was elevated by cascading multiple resonators \cite{Yang2020}. For the device in this work, the trade off is that the NRIR window is transported to higher input powers, and that the footprint of the device is also increased. We remind that $\alpha_n=\alpha/2\kappa=\alpha L_c/pi$, where $L_c$ is the coupling length of the coupler. Also note that since $\alpha_n$ is directly proportional to the coupling length, we can attain any value of $\alpha_n$ just by increasing the coupling length of the coupler. This offers some design flexibility and freedom in the choice of the SA nonlinear material. A material theoretically and experimentally proven to offer broadband SA in the NIR spectral region is graphene, which is an ideal candidate for an integrated physical implementation of this work \cite{Chatzidimitriou:20, Pitilakis_Chatzi2020}.

Furthermore, a nonlinear nonreciprocal directional coupler should inherit the considerable bandwidth (tens of nanometers) of the directional coupler as long as the SA effect is also relatively broadband. To get a qualitative feeling of the broadband operation lets consider a simple example: assume that the combined effect of the coupling length dispersion and material absorption dispersion is summarized to $\pm10\%$ of $\alpha_n=15$. From Fig.~\ref{fig:numerical}(a) and (b) we can observe that the coupler would still exhibit almost the same NRIR with a slight change in the average input power and ILs. The broadband operation, can be considered as the main advantage that such a device has over resonant nonlinear nonreciprocal devices \cite{Sounas:2018, Yoon2018}. Also, due to the high bandwidth, fabrication-related deviations are more smooth and forgiving in a directional coupler than in a cavity. On a final note, the main limiting factor for the device described here should be any linear non-saturable losses, which, depending on the length of the coupler, could greatly increase the ILs.

\section{Conclusions} \label{sec:5:conclusions}

Our theoretical analysis of a passive non-Hermitian photonic coupler with saturable losses proves that such a device can act as a promising nonlinear nonreciprocal element. Analytical and semi-analytical relations extracted for the NRIR highlight the role and importance of the EPs in the operation of the system, while also providing physical insight into the non-Hermitian dynamics that drive the system: In the forward direction transmission is gradually increased with increasing input power as the exit port of the coupler coincides with lossless waveguide and the low-loss eigenmode. On the other hand, in the backward direction, the exit port coincides with the lossy waveguide and thus input power coupled from the lossless to the lossy waveguide cannot exit the device unless it sufficiently saturates absorption. Also, in the backward direction the more power that is initially coupled to the nonlinear waveguide the easier it becomes for the coupling to continue so that backward transmission demonstrates a very sharp increase after an input power threshold. The sharpness of the increase is related to the contrast of losses between the eigenmodes. Furthermore, it was proven that the NRIR of the nonlinear coupler could be increased without sacrificing transmission, by just decreasing the coupling coefficient (and increasing the coupling length). The theory developed was confirmed by numerical simulations, which were also used to extract important performance metrics for practical applications. The results showed nonreciprocal behaviour with acceptable average forward ILs, large NRIR and potentially highly broadband operation due to the non-resonant nature of the system.

\begin{acknowledgments}
The research work was supported by the Hellenic Foundation for Research and Innovation (H.F.R.I.) under the ``First Call for H.F.R.I. Research Projects to support Faculty members and Researchers and the procurement of high-cost research equipment grant.'' (Project Number: HFRI-FM17-2086)
\end{acknowledgments}

%

\end{document}